\documentstyle{cupconf}
\input psfig

\ifoldfss
\else
  \ifnfssone
    \newmathalphabet{\mathit}
      \addtoversion{normal}{\mathit}{cmr}{m}{it}
      \addtoversion{bold}{\mathit}{cmr}{bx}{it}
    \newmathalphabet{\mathcal}
      \addtoversion{normal}{\mathcal}{cmsy}{m}{n}
    \else
    \ifnfsstwo
    \fi
  \fi
\fi

\def\hexnumber#1{\ifcase#1 0\or1\or2\or3\or4\or5\or6\or7\or8\or9\or
 A\or B\or C\or D\or E\or F\fi }

\makeatletter
\ifx\CUP@mtlplain@loaded\undefined
\else
\fi
\makeatother
 \makeatletter
 \ifx\CUP@mtlplain@loaded\undefined
   \font\tenbmi=cmmib10 at 10pt
   \font\sevenbmi=cmmib10 at 7pt
   \font\fivebmi=cmmib10 at 5pt

   \newfam\bmifam
   \textfont\bmifam=\tenbmi
   \scriptfont\bmifam=\sevenbmi
   \scriptscriptfont\bmifam=\fivebmi
   
 \fi
 \makeatother
\ifnfsstwo

\fi
\ifnfssone

\fi
\ifoldfss

\fi

\mathchardef\varLambda="0103

\makeatletter
\ifx\CUP@mtlplain@loaded\undefined
\else
\fi
\makeatother
\makeatletter
\ifx\CUP@mtlplain@loaded\undefined
  \font\tenbms=cmbsy10
  \font\sevenbms=cmbsy10 at 7pt
  \font\fivebms=cmbsy10 at 5pt
  \newfam\bmsfam
  \textfont\bmsfam=\tenbms
  \scriptfont\bmsfam=\sevenbms
  \scriptscriptfont\bmsfam=\fivebms

  \edef\bsy@{\hexnumber\bmsfam}
  \mathchardef\bnabla="0\bsy@72
\fi
\makeatother
\title[EVN Observations of GRS1915+105]{EVN Observations of GRS1915+105}

\author[L. Feretti {\it et al.\/}]
{L. FERETTI$^1$, G. GIOVANNINI$^1$, M. TORDI$^1$, T. VENTURI$^1$,\\ 
S. MASSAGLIA$^2$, G. BODO$^3$,  E. TRUSSONI$^3$, M. GLIOZZI$^4$,\\ 
M. TAVANI$^5$, J. CONWAY$^6$, 
A. FOLEY$^7$, D. GRAHAM$^8$, \\ A. KUS$^9$, R. SPENCER$^{10}$, 
C. TRIGILIO$^{11}$
}
 
\affiliation{
$^1$IRA Bologna, $^2$Univ. Torino, $^3$Oss. Torino,  $^4$MPE Garching, 
$^5$IFCTR Milano,\\  $^6$Onsala Observat., $^7$ NFRA Dwingeloo, 
$^8$ MPifR Bonn,  $^9$ Torun Observ.,\\ $^{10}$NRAL Jodrell Bank,  
$^{11}$ IRA Noto  }

\begin{document}
\ifnfssone
\else
  \ifnfsstwo
  \else
    \ifoldfss
      \let\mathcal\cal
      \let\mathrm\rm
      \let\mathsf\sf
    \fi
  \fi
\fi

\maketitle

\begin{abstract}
\end{abstract}

\firstsection 
\section{Introduction}

The source GRS1915+105 is an X-ray transient source, at the distance of 
11 kpc, which represents the first case of apparent 
superluminal motion in a source  within our own Galaxy (Mirabel and 
Rodriguez 1994, Fender et al. 1999). 
The source, constantly monitored in radio at the Green Bank Interferometer
(GBI) in Virginia, USA, shows quiescent periods, with radio flux densities
of a few mJy, periods of radio activity, with fluxes around 200-300 mJy 
at 2.2 GHz, and rapid major flares, up to 1 Jy or more. 

\section{Observations and Results}

We obtained EVN observations aimed at observing the source 
in a radio-loud state (Fig. 1). TOO observations were performed during gaps of 
the official observing schedule on 31/05, 02/06 and 09/06 (1998)
at 6 cm, 6 cm and 18 cm, respectively. Integration times were of about 
4 h in each observation.

\begin{figure}[here] 
\centerline{\psfig{figure=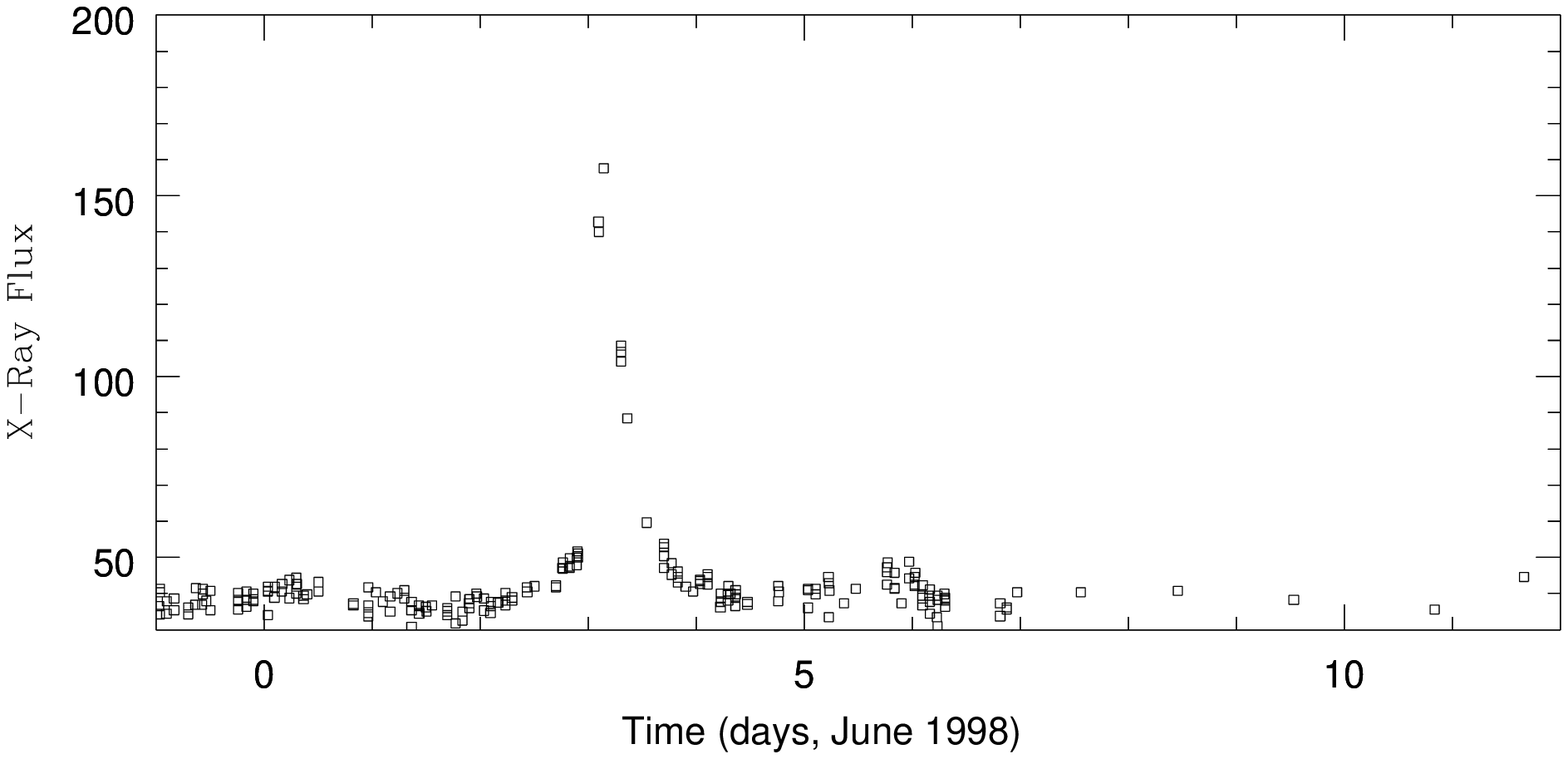,height=4.0truecm}
\psfig{figure=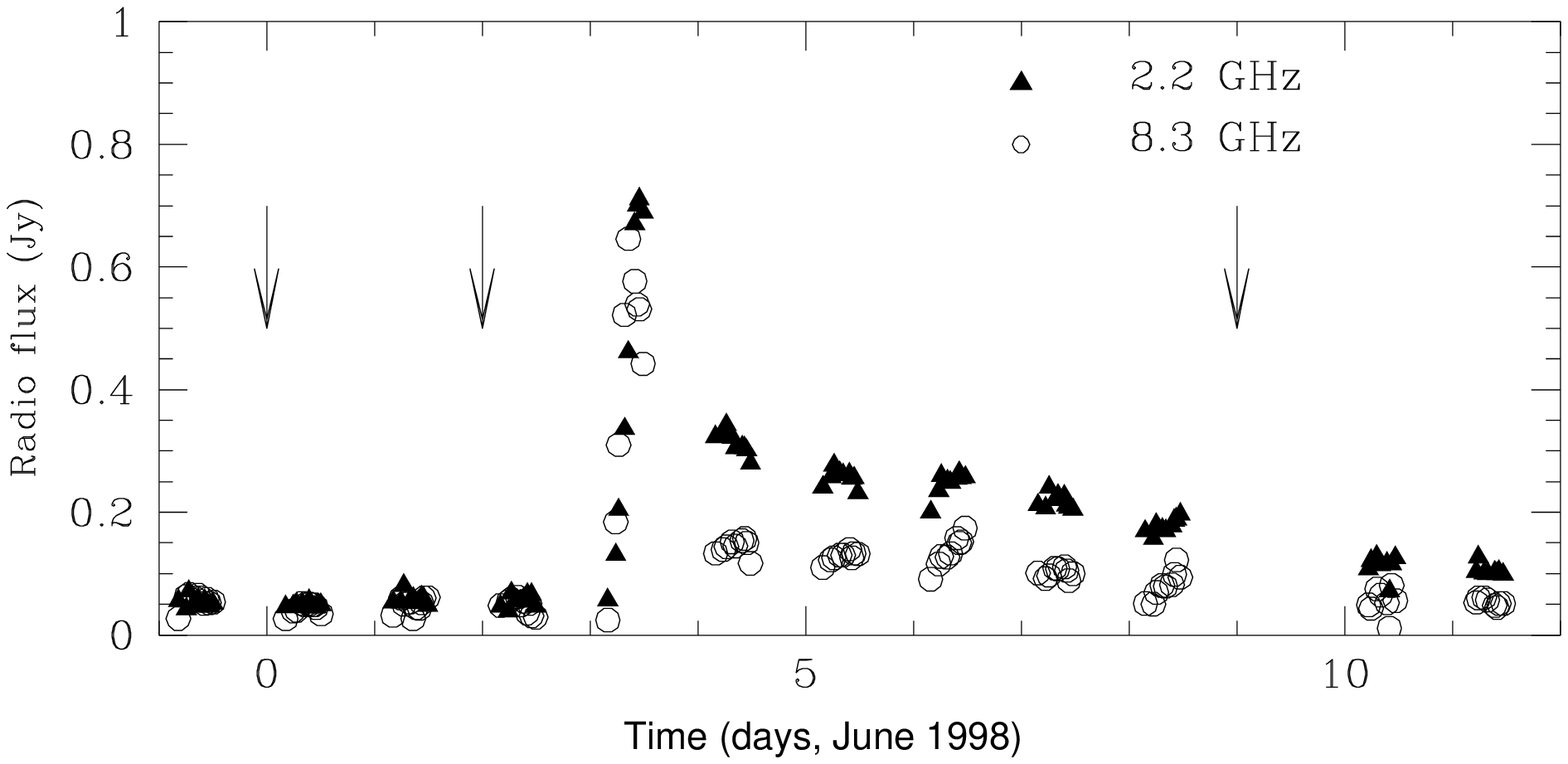,height=4.0truecm}}
  \caption{
Source light curve in X-rays (XTE, left) and in radio (GBI, right), with
arrows indicating the epochs of the EVN observations.
} 
\end{figure}

The telescopes involved in the observations were  Medicina, Noto, 
Effelsberg, Jodrell, Onsala,  Westerbork and Torun.
The data  have been calibrated at JIVE (Dwingeloo).
Source images (Fig. 2) have been produced with the AIPS package. 

\begin{figure} 
\centerline {\psfig{figure=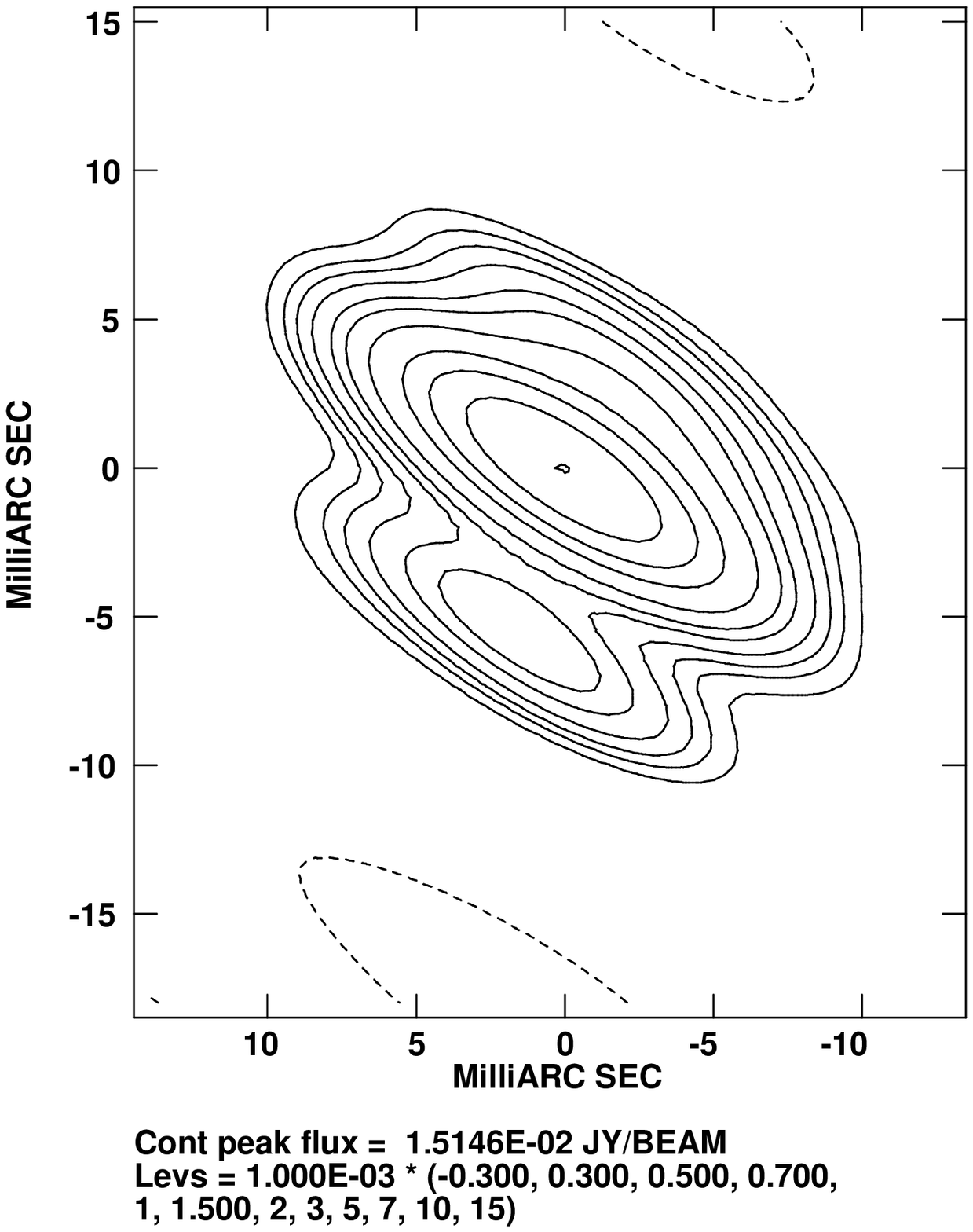,height=6.0truecm}
\psfig{figure=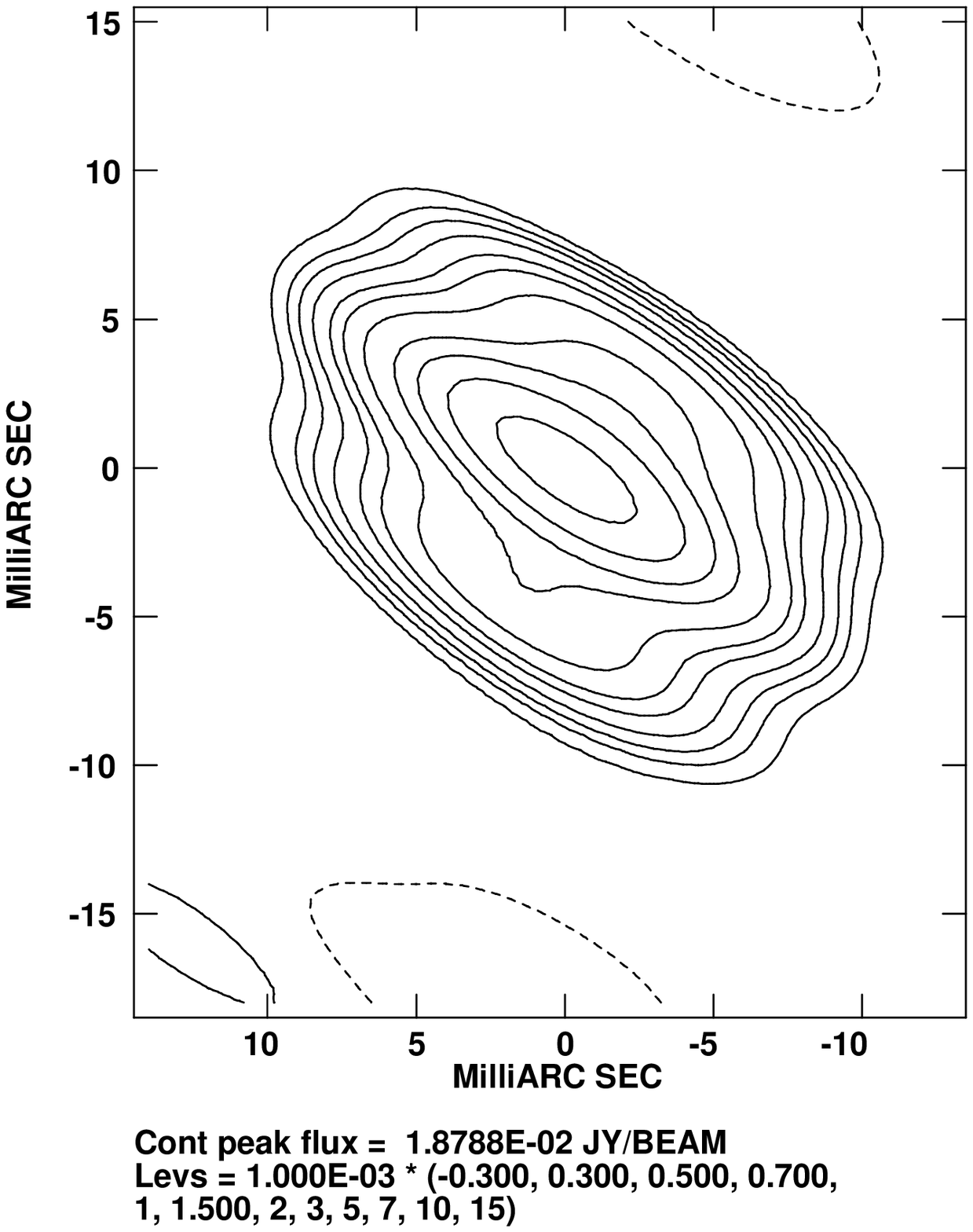,height=6.0truecm}
\psfig{figure=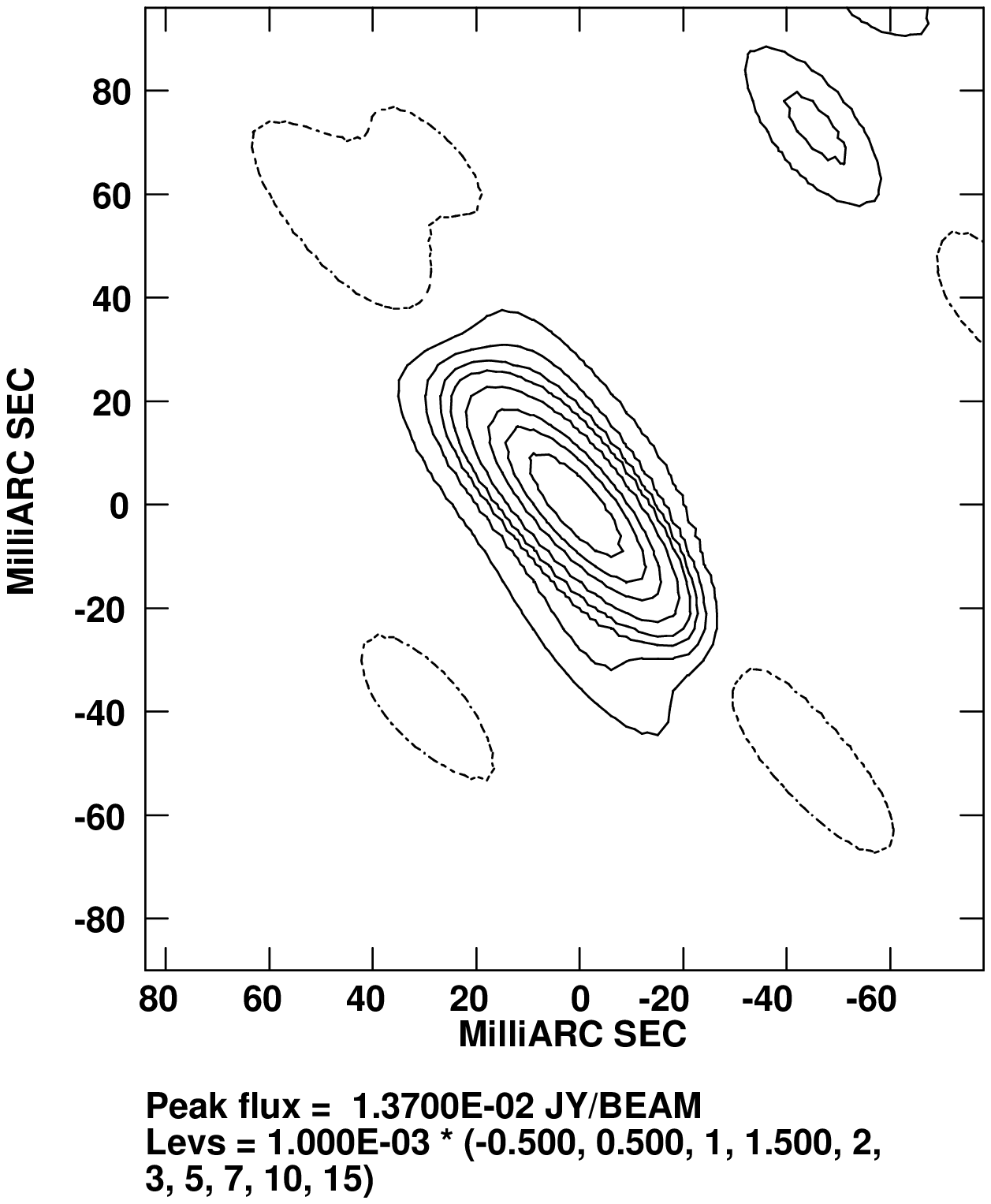,height=6.0truecm}}
  \caption{Images of the 3 EVN observations.  The first contour is 
always at the 2$\sigma$ noise level.
Left panel: May 31 1998, 6 cm,
HPBW = 10$\times$ 3 mas, Central panel: June 2 1998, 6 cm,
HPBW = 10$\times$ 3 mas, Right panel: June 9 1998, 18 cm,
HPBW = 39$\times$ 12 mas 
} 

\end{figure}

In the two images obtained before the major flare, the source
is resolved, showing two opposite asymmetric jets. The main jet is in the SE
direction in position angle $\sim$160$^{\circ}$, 
consistent with the orientation of the main jet present 
in previous MERLIN and VLA data (Fender et al. 1999, Mirabel 
\& Rodriguez 1994). 
The total source extent is of $\sim$13 mas,
corresponding to 2.1 $\times$10$^{15}$ cm ($\sim$140 AU). 
The total flux shows a slight increase (from 22 mJy to 31 mJy)
between the two epochs. These  flux  measurements are 
roughly consistent with the GBI fluxes.

In the third observation, obtained  after the flare,
the source is unresolved.
The total flux is 17 mJy, 
much lower than the GBI flux. This would imply the presence
of extended structure  
not detected in the present observation, supporting the idea 
that during the flare strong high  velocity components have been ejected.

\section{Discussion and Conclusions}

From the present results, it is evident that steady jets exist in the
source also before a major flare. This confirms the findings of Dhawan 
et al. (2000), but the jets detected in the present case are much more
extended. 

Assuming that the jets are relativistic and intrinsically symmetric,
we can derive an estimate of the velocity from the jet to counter-jet 
brightness ratio and from the jet-length ratio. Using a source orientation
to the plane of the sky of 66$^{\circ}$ (Fender et al. 1999), 
we derive a range for the velocity of 0.2-0.6c up to
a distance of $\sim$8 mas ($\sim$ 90 AU).
With this velocity, we would expect to detect  proper motion 
 between  the two images of May 31 and June 2. This is not observed, 
the source structure being rather stable. 
The absence of proper motion 
could be due to a continuous ejection in a stable jet. 

By comparing the present velocity estimate with the 
superluminal velocity measured after a major flare, we could suggest that the
flare is producing a strong ejection at much higher speed.
The high resolution image obtained by Dhawan et al (2000)
before a major flare, shows a quite symmetric source on a scale of 
$\sim$50 AU, with a jet velocity of $\sim$0.1c. The result obtained
from the present data  would imply that 
the jet accelerates at distance between 50 AU and 90 AU from the black hole. 

\begin{acknowledgments}
This work has been partially supported by the
European Commission's TMR-LSF programme under contract ERBFMGECT950012. 

\end{acknowledgments}

\end{document}